# The Dimensions of Individual Strings and Sequences

Jack H. Lutz [*]
Department of Computer Science
Iowa State University
Ames, IA 50011, USA
lutz@cs.iastate.edu

**Abstract**

A constructive version of Hausdorff dimension is developed using constructive *supergales*, which are betting strategies that generalize the constructive supermartingales used in the theory of individual random sequences. This constructive dimension is used to assign every individual (infinite, binary) sequence $S$ a *dimension*, which is a real number $\dim(S)$ in the interval $[0, 1]$. Sequences that are random (in the sense of Martin-Löf) have dimension 1, while sequences that are decidable, $\Sigma_1^0$, or $\Pi_1^0$ have dimension 0. It is shown that for every $\Delta_2^0$-computable real number $\alpha$ in $[0,1]$ there is a $\Delta_2^0$ sequence $S$ such that $\dim(S) = \alpha$.

A discrete version of constructive dimension is also developed using *termgales*, which are supergale-like functions that bet on the terminations of (finite, binary) strings as well as on their successive bits. This discrete dimension is used to assign each individual string $w$ a *dimension*, which is a nonnegative real number $\dim(w)$. The dimension of a sequence is shown to be the limit infimum of the dimensions of its prefixes.

The Kolmogorov complexity of a string is proven to be the product of its length and its dimension. This gives a new characterization of algorithmic information and a new proof of Mayordomo's recent theorem stating that the dimension of a sequence is the limit infimum of the average Kolmogorov complexity of its first $n$ bits.

Every sequence that is random relative to any computable sequence of coin-toss biases that converge to a real number $\beta$ in $(0, 1)$ is shown to have dimension $\mathcal{H}(\beta)$, the binary entropy of $\beta$.

**Keywords**: algorithmic information, computability, constructive dimension, dimension, entropy, gales, Hausdorff dimension, Kolmogorov complexity, Kullback-Leibler divergence, martingales, randomness, supergales, termgales.

## 1 Introduction

One of the most dramatic achievements of the theory of computing was Martin-Löf's 1966 use of constructive measure theory to give the first satisfactory definition of the randomness of individual infinite binary sequences [26]. The search for such a definition had been a major object of early twentieth-century research on the foundations of probability, but a rigorous mathematical formulation had proven so elusive that the search had been all but abandoned more than two decades earlier. Martin-Löf's definition says precisely which infinite binary sequences are random and which are not. The definition is probabilistically convincing in that it requires each random sequence to pass every algorithmically implementable statistical test of randomness. The definition is also robust in that subsequent definitions by Schnorr [36, 37, 38], Levin [20], Chaitin [6], Solovay [44], and Shen' [40, 41], using a variety of different approaches, all define exactly the same sequences to be random. It is noteworthy that all these approaches, like Martin-Löf's, make essential use of the theory of computing.

---
[*]This work was supported in part by National Science Foundation Grants 9610461 and 9988483.

One useful characterization of random sequences is that they are those sequences that have maximal algorithmic information content. Specifically, if $K(S[0..n-1])$ denotes the Kolmogorov complexity (algorithmic information content) of the first $n$ bits of an infinite binary sequence $S$, then Levin [20] and Chaitin [6] have shown that $S$ is random if and only if there is a constant $c$ such that for all $n$, $K(S[0..n-1]) \geq n - c$. Indeed Kolmogorov [17] developed what is now called $C(x)$, the "plain Kolmogorov complexity," in order to formulate such a definition of randomness, and Martin-Löf, who was then visiting Kolmogorov, was motivated by this idea when he defined randomness. (The quantity $C(x)$ was also developed independently by Solomonoff [43] and Chaitin [4, 5].) Martin-Löf [27] subsequently proved that $C(x)$ cannot be used to characterize randomness, and Levin [20] and Chaitin [6] introduced a technical modification of $C(x)$, now called $K(x)$, the "Kolmogorov complexity," in order to prove the above characterization of random sequences. Schnorr [38] proved a similar characterization in terms of another variant, called the "monotone Kolmogorov complexity."

One conclusion to be drawn from these characterizations is that the definition of random sequences distinguishes those sequences that have maximal algorithmic information content from those that do not. It offers no quantitative classification of the sequences that have less than maximal algorithmic information content. From a technical point of view, this aspect of the definition arises from its use of constructive measure, which is an algorithmic effectivization of classical Lebesgue measure. Specifically, an infinite binary sequence $S$ is random if the singleton set $\{S\}$ does not have constructive measure 0, and is nonrandom if $\{S\}$ does have constructive measure 0. Neither Lebesgue measure nor constructive measure offers quantitative distinctions among measure 0 sets.

In 1919, Hausdorff [13] augmented classical Lebesgue measure theory with a theory of dimension. This theory assigns to every subset $X$ of a given metric space a real number $\dim_H(X)$, which is now called the *Hausdorff dimension* of $X$. In this paper we are interested in the case where the metric space is the Cantor space $\mathbf{C}$, consisting of all infinite binary sequences. In this case, the Hausdorff dimension of a set $X \subseteq \mathbf{C}$ (which is defined precisely in section 3 below) is a real number $\dim_H(X) \in [0,1]$. The Hausdorff dimension is monotone, with $\dim_H(\emptyset) = 0$ and $\dim_H(\mathbf{C}) = 1$. Moreover, if $\dim_H(X) < \dim_H(\mathbf{C})$, then $X$ is a measure 0 subset of $\mathbf{C}$. Hausdorff dimension thus offers a quantitative classification of measure 0 sets. Moreover, Ryabko [33, 34, 35] Staiger [46, 47], and Cai and Hartmanis [3] have all proven results establishing quantitative relationships between Hausdorff dimension and Kolmogorov complexity.

Just as Hausdorff [13] augmented Lebesgue measure with a theory of dimension, this paper augments the theory of individual random sequences with a theory of the dimensions of individual sequences. Specifically, we develop a constructive version of Hausdorff dimension and use this to assign every sequence $S \in \mathbf{C}$ a *dimension* $\dim(S) \in [0,1]$. Sequences that are random have dimension 1, while sequences that are decidable, $\Sigma_1^0$, or $\Pi_1^0$ have dimension 0. For every real number $\alpha \in [0,1]$ there is a sequence $S$ such that $\dim(S) = \alpha$. Moreover, if $\alpha$ is $\Delta_2^0$-computable, then there is a $\Delta_2^0$ sequence $S$ such that $\dim(S) = \alpha$. (This generalizes the well-known existence of $\Delta_2^0$ sequences that are random.)

Our development of constructive dimension is based on *supergales*, which are natural generalizations of the constructive supermartingales used by Schnorr [36, 37, 38] to characterize randomness. In a recent paper [25] we have shown that supergales can be used to characterize the classical Hausdorff dimension, and that resource-bounded supergales can be used to define dimension in complexity classes. In the present paper we use constructive (lower semicomputable) supergales to develop constructive dimension. The dimension of a sequence $S \in \mathbf{C}$ is then the constructive dimension of the singleton set $\{S\}$. Constructive dimension differs markedly from both classical Hausdorff dimension and the resource-bounded dimension developed in [25], primarily due to the existence of supergales that are optimal. These optimal supergales are analogous to universal tests of randomness in the theory of random sequences.

Supergales, like supermartingales, are strategies for betting on the successive bits of infinite binary sequences. In order to define the dimensions of individual strings $w \in \{0,1\}^*$, we introduce *termgales*, which are supergale-like functions that bet on the terminations of strings as well as on their successive bits. Using termgales, we assign each binary string $w$ a *dimension* $\dim(w)$, which is a nonnegative real number. We show that for every sequence $S \in \mathbf{C}$,

$$\dim(S) = \liminf_{n \to \infty} \dim(S[0..n-1]). \tag{1.1}$$

We use dimension to prove a new characterization of Kolmogorov complexity. Specifically, we show that there is a constant $c$ such that for all $w \in \{0,1\}^*$,

$$\Big| K(w) - |w| \dim(w) \Big| \leq c. \tag{1.2}$$

That is, the Kolmogorov complexity of a string is (to within a constant additive term) the product of the string's length and its dimension. This characterization of Kolmogorov complexity in terms of a constructivized, discretized version of Hausdorff's 1919 theory of dimension is reminiscent of (and technically related to) the well-known characterization by Levin [20, 21] and Chaitin [6] of Kolmogorov complexity in terms of constructivized discrete probability, i.e., the fact that there is a constant $c' \in \mathbb{N}$ such that for all $w \in \{0,1\}^*$,

$$\Big| K(w) - \log \frac{1}{\mathbf{m}(w)} \Big| \leq c', \tag{1.3}$$

where $\mathbf{m}$ is an optimal constructive subprobability measure on $\{0,1\}^*$.

Taken together, (1.1) and (1.2) provide a new proof of Mayordomo's recent theorem [28] stating that for every sequence $S \in \mathbf{C}$,

$$\dim(S) = \liminf_{n \to \infty} \frac{K(S[0..n-1])}{n}. \tag{1.4}$$

Facts (1.2) and (1.4) justify the intuition that the dimension of a string or sequence is a measure of its *algorithmic information density*.

We also investigate the dimensions of sequences that are random relative to computable sequences of convergent coin-toss biases. Specifically, let $\vec{\beta} = (\beta_0, \beta_1, \beta_2, \ldots)$ be any computable sequence of real numbers $\beta_i \in [0,1]$ that converge to a real number $\beta \in (0,1)$ (which must therefore be $\Delta_2^0$-computable). We show that if $R$ is any sequence in $\mathbf{C}$ that is random with respect to $\vec{\beta}$ (i.e., a random outcome of a random experiment in which for each $i$, independently of all other $j$, the $i^{\text{th}}$ bit of $R$ is decided by tossing a 0/1-valued coin whose probability of 1 is $\beta_i$), then the dimension of $R$ is $\mathcal{H}(\beta)$, the binary Shannon entropy of $\beta$.

We defer discussion of some significant related work until late in the paper, where more context is available. Specifically, results by Schnorr [37, 39], Ryabko [32, 33, 34, 35], Staiger [46, 47, 45], and Cai and Hartmanis [3] that relate martingales, supermartingales, and Kolmogorov complexity to Hausdorff dimension are discussed at the end of section 6. Classical work by Besicovitch [1], Good [12], and Eggleston [9] relating limiting frequencies and Shannon entropy to Hausdorff dimension is described briefly in section 7.

## 2 Preliminaries

We use the set $\mathbb{Z}$ of integers, the set $\mathbb{Z}^+$ of (strictly) positive integers, the set $\mathbb{N}$ of natural numbers (i.e., nonnegative integers), the set $\mathbb{Q}$ of rational numbers, the set $\mathbb{R}$ of real numbers, and the set $[0, \infty)$ of nonnegative reals.

A *string* is a finite, binary string $w \in \{0,1\}^*$. We write $|w|$ for the length of a string $w$ and $\lambda$ for the empty string. For $i, j \in \{0, \ldots, |w|-1\}$, we write $w[i..j]$ for the string consisting of the $i^{\text{th}}$ through the $j^{\text{th}}$ bits of $w$ and $w[i]$ for $w[i..i]$, the $i^{\text{th}}$ bit of $w$. Note that the $0^{\text{th}}$ bit $w[0]$ is the leftmost bit of $w$ and that $w[i..j] = \lambda$ if $i > j$. A *sequence* is an infinite, binary sequence. If $S$ is a sequence and $i, j \in \mathbb{N}$, then the notations $S[i..j]$ and $S[i]$ are defined exactly as for strings. We work in the *Cantor space* $\mathbf{C}$ consisting of all sequences. A string $w \in \{0,1\}^*$ is a *prefix* of a sequence $S \in \mathbf{C}$, and we write $w \sqsubseteq S$, if $S[0..|w|-1] = w$. The *cylinder generated by* a string $w \in \{0,1\}^*$ is $\mathbf{C}_w = \{S \in \mathbf{C} | w \sqsubseteq S\}$. Note that $\mathbf{C}_\lambda = \mathbf{C}$.

We also make passing references to $\Pi_1^0$, $\Delta_2^0$, and $\Sigma_2^0$ sets of sequences. These refer to the arithmetical (i.e., effective Borel) hierarchy of sets of sequences and are not central to our development. The interested reader is referred to [31] or [29] for discussion of this hierarchy.

The *support* of a sequence $S \in \mathbf{C}$ is

$$\text{supp}(S) = \{n \in \mathbb{N} | S[n] = 1\}.$$

The arithmetical hierarchy of sequences is defined from the arithmetical hierarchy of subsets of $\mathbb{N}$ using the support function. Thus, for example, a sequence $S \in \mathbf{C}$ is *computably enumerable*, and we write $S \in \Sigma_1^0$, if $\text{supp}(S)$ is a $\Sigma_1^0$ (i.e., computably enumerable) subset of $\mathbb{N}$. Similarly, $S \in \Pi_1^0$ if $\mathbb{N} - \text{supp}(S)$ is computably enumerable, and $S \in \Delta_2^0$ if $\text{supp}(S)$ is decidable relative to the halting oracle.

If $S, T \in \mathbf{C}$, then $S$ is *1-reducible* to $T$, and we write $S \leq_1 T$, if there is a computable, one-to-one function $f : \mathbb{N} \to \mathbb{N}$ such that for all $n \in \mathbb{N}$, $S[n] = T[f(n)]$. The sequences $S$ and $T$ are *1-equivalent*, and we write $S \equiv_1 T$, if $S \leq_1 T$ and $T \leq_1 S$.

**Definition.** A *subprobability measure* on $\{0,1\}^*$ is a function $p : \{0,1\}^* \to [0,1]$ such that

$$\sum_{w \in \{0,1\}^*} p(w) \leq 1. \tag{2.1}$$

A *probability measure* on $\{0,1\}^*$ is a subprobability measure on $\{0,1\}^*$ that satisfies (2.1) with equality. A *subprobability supermeasure* on the Cantor space $\mathbf{C}$ is a function $\nu : \{0,1\}^* \to [0,1]$ such that

$$\nu(\lambda) \leq 1 \tag{2.2}$$

and for all $w \in \{0,1\}^*$,

$$\nu(w) \geq \nu(w0) + \nu(w1). \tag{2.3}$$

A *subprobability measure* on $\mathbf{C}$ is a subprobability supermeasure on $\mathbf{C}$ that satisfies (2.3) with equality for all $w \in \{0,1\}^*$, and a *probability measure* on $\mathbf{C}$ is a subprobability measure on $\mathbf{C}$ that satisfies (2.2) with equality. Intuitively, if $\nu$ is a probability measure on $\mathbf{C}$ and $w \in \{0,1\}^*$, then $\nu(w)$ is the probability that $w \sqsubseteq S$ when the sequence $S \in \mathbf{C}$ is "chosen according to the probability measure $\nu$."

A *bias* is a real number $\beta \in [0,1]$. Intuitively, if we toss a 0/1-valued coin with bias $\beta$, then $\beta$ is the probability of the outcome 1. A *bias sequence* is a sequence $\vec{\beta} = (\beta_0, \beta_1, \beta_2, \dots)$ of biases. If $\vec{\beta}$ is a bias sequence, then the $\vec{\beta}$-*coin-toss probability measure* is the probability $\mu^{\vec{\beta}}$ on $\mathbf{C}$ defined by

$$\mu^{\vec{\beta}}(w) = \prod_{i=0}^{|w|-1} \beta_i(w), \tag{2.4}$$

where $\beta_i(w) = (2\beta_i - 1)w[i] + (1 - \beta_i)$, i.e., $\beta_i(w) = \textbf{if } w[i] \textbf{ then } \beta_i \textbf{ else } 1 - \beta_i$. That is, $\mu^{\vec{\beta}}$ is the probability that $S \in \mathbf{C}_w$ when $S \in \mathbf{C}$ is chosen according to a random experiment in which for each $i$, independently of all other $j$, the $i^{\text{th}}$ bit of $S$ is decided by tossing a 0/1-valued coin whose probability of 1 is $\beta_i$. In the case where the biases $\beta_i$ are all the same, i.e., $\vec{\beta} = (\beta, \beta, \beta, \dots)$ for some $\beta \in [0,1]$, we write $\mu^\beta$ for $\mu^{\vec{\beta}}$, and (2.4) simplifies to

$$\mu^\beta(w) = (1-\beta)^{\#(0,w)} \beta^{\#(1,w)}, \tag{2.5}$$

where $\#(b, w)$ is the number of times the bit $b$ appears in the string $w$. The *uniform probability measure* on $\mathbf{C}$ is the probability measure $\mu = \mu^{\frac{1}{2}}$, for which (2.5) simplifies to

$$\mu(w) = 2^{-|w|} \tag{2.6}$$

for all $w \in \{0,1\}^*$.

We use several conditions involving the computability of real numbers and real-valued functions in this paper.

**Definition.** Let $f : D \to \mathbb{R}$, where $D$ is some discrete domain such as $\mathbb{N}, \{0,1\}^*, \mathbb{N} \times \{0,1\}^*$, etc.

1. $f$ is *computable* if there is a computable function $\hat{f} : D \times \mathbb{N} \to \mathbb{Q}$ such that for all $(x, r) \in D \times \mathbb{N}$, $|\hat{f}(x, r) - f(x)| \leq 2^{-r}$.

2. $f$ is *lower semicomputable* if there is a computable function $\hat{f} : D \times \mathbb{N} \to \mathbb{Q}$ such that
   (a) for all $(x, t) \in D \times \mathbb{N}$, $\hat{f}(x, t) \leq \hat{f}(x, t+1) < f(x)$, and
   (b) for all $x \in D$, $\lim_{t \to \infty} \hat{f}(x, t) = f(x)$.

3. $f$ is $\Delta_2^0$-*computable* if $f$ is computable relative to the halting oracle.

The following facts are well known and easy to verify.

(i) Computability implies lower semicomputability, lower semicomputability implies $\Delta_2^0$-computability, and the converses of these statements do not hold.

(ii) A function $f : D \to \mathbb{R}$ is computable if and only if the functions $f$ and $-f$ are both lower semicomputable.

(iii) A function $f : D \to \mathbb{R}$ is lower semicomputable if and only if its *lower graph*

$$\text{Graph}^-(f) = \{(x, s) \in D \times \mathbb{Q} | s < f(x)\}$$

is computably enumerable.

A sequence $\vec{\alpha} = (\alpha_0, \alpha_1, \alpha_2, \dots)$ of real numbers is *computable* if the associated function $f_{\vec{\alpha}} : \mathbb{N} \to \mathbb{R}$, defined by $f_{\vec{\alpha}}(i) = \alpha_i$ for all $i \in \mathbb{N}$, is computable. A real number $\alpha$ is *computable* if the sequence $\vec{\alpha} = (\alpha, \alpha, \alpha, \dots)$ is computable. Note that if $\vec{\beta}$ is a computable bias sequence, then $\mu^{\vec{\beta}}$ is a computable probability measure on $\mathbf{C}$.

**Definition.** A subprobability measure on $\{0,1\}^*$ or a subprobability supermeasure on $\mathbf{C}$ is *constructive* if it is lower semicomputable.

**Definition.** If $\mathcal{F}$ is a class of functions from $\{0,1\}^*$ into $[0,\infty)$, then an *optimal* element of $\mathcal{F}$ is a function $g \in \mathcal{F}$ such that for every $f \in \mathcal{F}$ there is a real constant $\alpha > 0$ such that for all $w \in \{0,1\}^*$, $g(w) \geq \alpha f(w)$.

The following theorem is one of the cornerstones of algorithmic information theory.

**Theorem 2.1.** (Levin [52])

1. There is an optimal constructive subprobability measure $\mathbf{m}$ on $\{0,1\}^*$.

2. There is an optimal constructive subprobability supermeasure $\mathbf{M}$ on $\mathbf{C}$.

Throughout this paper we fix $\mathbf{m}$ and $\mathbf{M}$ as in Theorem 2.1.

The reader is referred to the text by Li and Vitanyi [22] for the definition and basic properties of the Kolmogorov complexity $K(w)$, defined for strings $w \in \{0,1\}^*$. The main property of Kolmogorov complexity that we use here is the following theorem, which is another cornerstone of algorithmic information theory.

**Theorem 2.2.** (Levin [20, 21], Chaitin [6]) *There is a constant $c \in \mathbb{N}$ such that for all $w \in \{0,1\}^*$,*

$$\left| K(w) - \log \frac{1}{\mathbf{m}(w)} \right| \leq c.$$

Given a set $A \subseteq \{0,1\}^*$ and $n \in \mathbb{N}$, we use the abbreviations $A_{=n} = A \cap \{0,1\}^n$ and $A_{\leq n} = A \cap \{0,1\}^{\leq n}$. A *prefix set* is a set $A \subseteq \{0,1\}^*$ such that no element of $A$ is a prefix of another element of $A$.

Let $X$ be a $k$-fold product of intervals, each of which is $(0,1)$ or $[0,1]$. If $g : X \to \mathbb{R}$ and $\vec{\alpha} = (\alpha_1, \ldots, \alpha_k) \in X$, then we sometimes use "$g(\vec{\alpha})$" as an abbreviation for the random variable $\xi : \{0,1\} \to \mathbb{R}$ defined by $\xi(1) = g(\alpha_1, \ldots, \alpha_k)$ and $\xi(0) = g(1-\alpha_1, \ldots, 1-\alpha_k)$. If $\beta \in [0,1]$, then we also use $\beta$ as an abbreviation for the probability measure $p$ on $\{0,1\}$ in which $p(1) = \beta$. Thus, for example,

$$\mathrm{E}_\beta g(\vec{\alpha}) = \beta g(\alpha_1, \ldots, \alpha_k) + (1-\beta) g(1-\alpha_1, \ldots, 1-\alpha_k)$$

is the expected value of "the random variable $g(\vec{\alpha})$" with respect to "the probability measure $\beta$." In particular, Shannon's *binary entropy* function $\mathcal{H} : [0,1] \to [0,1]$ is defined by

$$\mathcal{H}(\beta) = \mathrm{E}_\beta \log \frac{1}{\beta},$$

with the proviso that $0 \log \frac{1}{0} = 0$ so that $\mathcal{H}$ is continuous on $[0,1]$. Similarly, the *binary Kullback-Leibler divergence* function $D : [0,1] \times (0,1) \to \mathbb{R}$ is defined by

$$D(\beta \parallel \alpha) = \mathrm{E}_\beta \log \frac{\beta}{\alpha},$$

with the proviso that $0 \log \frac{0}{p} = 0$ so that $D$ is continuous on $[0,1] \times (0,1)$. It is well-known that $D(\beta \parallel \alpha) \geq 0$, with equality if and only if $\beta = \alpha$. See the text by Cover and Thomas [8] for further discussion of $\mathcal{H}(\beta)$ and $D(\beta \parallel \alpha)$.

Falconer [10] provides a good overview of Hausdorff dimension.

## 3 Gales and Constructive Dimension

In this section we define gales and supergales and use these to define classical and constructive Hausdorff dimensions in the Cantor space **C**. Our definitions are slightly more general than those in [25] because here we need to define gales and supergales relative to an arbitrary (not necessarily uniform) probability measure on **C**.

**Definition.** Let $\nu$ be a probability measure on **C**, and let $s \in [0, \infty)$.

1. A *$\nu$-s-supergale* is a function $d : \{0,1\}^* \to [0, \infty)$ that satisfies the condition

$$d(w)\nu(w)^s \geq d(w0)\nu(w0)^s + d(w1)\nu(w1)^s \tag{3.1}$$

   for all $w \in \{0,1\}^*$.

2. A *$\nu$-s-gale* is a $\nu$-s-supergale that satisfies (3.1) with equality for all $w \in \{0,1\}^*$.

3. A *$\nu$-supermartingale* is a $\nu$-1-supergale.

4. A *$\nu$-martingale* is a $\nu$-1-gale.

5. An *s-supergale* is a $\mu$-s-supergale.

6. An *s-gale* is a $\mu$-s-gale.

7. A *supermartingale* is a 1-supergale.

8. A *martingale* is a 1-gale.

**Observations 3.1.**   *1. A subprobability supermeasure on **C** is a 0-supergale $d$ with $d(\lambda) \leq 1$.*

*2. A subprobability measure on **C** is a 0-gale $d$ with $d(\lambda) \leq 1$.*

*3. A probability measure on **C** is a 0-gale $d$ with $d(\lambda) = 1$.*

The following obvious but useful observation shows how gales and supergales are affected by variation of the parameter $s$.

**Observation 3.2.** *Let $\nu$ be a probability measure on **C**, let $s, s' \in [0, \infty)$, and let $d, d' : \{0,1\}^* \to [0, \infty)$. Assume that*

$$d(w)\nu(w)^s = d'(w)\nu(w)^{s'}$$

*for all $w \in \{0,1\}^*$.*

*1. $d$ is a $\nu$-s-supergale if and only if $d'$ is a $\nu$-s'-supergale.*

*2. $d$ is a $\nu$-s-gale if and only if $d'$ is a $\nu$-s'-gale.*

For example, Observation 3.2 implies that a function $d : \{0,1\}^* \to [0, \infty)$ is an s-gale if and only if the function $d' : \{0,1\}^* \to [0, \infty)$ defined by $d'(w) = 2^{(1-s)|w|}d(w)$ is a martingale.

Our next lemma is a generalization of Kraft's inequality [8].

**Lemma 3.3.** *Let $d$ be a $\nu$-$s$-supergale, where $\nu$ is a probability measure on $\mathbf{C}$ and $s \in [0, \infty)$. Then for all $w \in \{0,1\}^*$ and all prefix sets $B \subseteq \{0,1\}^*$,*

$$\sum_{u \in B} d(wu)\nu(wu)^s \leq d(w)\nu(w)^s.$$

*Proof.* We first use induction on $n$ to show that the lemma holds for all prefix sets $B \subseteq \{0,1\}^{\leq n}$. For $n = 0$ this is trivial. Assume that it holds for $n$, and let $A \subseteq \{0,1\}^{\leq n+1}$ be a prefix set. Define the set

$$A' = \{u \in \{0,1\}^n \mid u0 \in A \text{ or } u1 \in A\},$$

and note that $A_{\leq n}$ is disjoint from $A'$. Note also that the set

$$B = A_{\leq n} \cup A'$$

is a prefix set. For all $w \in \{0,1\}^*$, we have

$$\begin{aligned}
\sum_{u \in A_{=n+1}} d(wu)\nu(wu)^s &\leq \sum_{u \in A'} [d(wu0)\nu(wu0)^s + d(wu1)\nu(wu1)^s] \\
&\leq \sum_{u \in A'} d(wu)\nu(wu)^s.
\end{aligned}$$

It follows by the induction hypothesis that for all $w \in \{0,1\}^*$,

$$\begin{aligned}
\sum_{u \in A} d(wu)\nu(wu)^s &= \sum_{u \in A_{\leq n}} d(wu)\nu(wu)^s + \sum_{u \in A_{=n+1}} d(wu)\nu(wu)^s \\
&\leq \sum_{u \in A_{\leq n}} d(wu)\nu(wu)^s + \sum_{u \in A'} d(wu)\nu(wu)^s \\
&= \sum_{u \in B} d(wu)\nu(wu)^s \\
&\leq d(w)\nu(w)^s.
\end{aligned}$$

This completes the proof that for all $n \in \mathbb{N}$ the lemma holds for all prefix sets $B \subseteq \{0,1\}^{\leq n}$.

To complete the proof of the lemma, let $B$ be an arbitrary prefix set. Then for all $w \in \{0,1\}^*$,

$$\sum_{u \in B} d(wu)\nu(wu)^s = \sup_{n \in \mathbb{N}} \sum_{u \in B_{\leq n}} d(wu)\nu(wu)^s \leq d(w)\nu(w)^s.$$

□

**Definition.** Let $d$ be a $\nu$-$s$-supergale, where $\nu$ is a probability measure on $\mathbf{C}$ and $s \in [0,\infty)$.

1. We say that $d$ *succeeds* on a sequence $S \in \mathbf{C}$ if $\limsup_{n \to \infty} d(S[0..n-1]) = \infty$.

2. The *success set* of $d$ is $S^\infty[d] = \{S \in \mathbf{C} \mid d \text{ succeeds on } S\}$.

We now show how to use the success sets of gales and supergales to define Hausdorff dimension.

**Notation.** Let $X \subseteq \mathbf{C}$.

1. $\mathcal{G}(X)$ is the set of all $s \in [0, \infty)$ such that there is an $s$-gale $d$ for which $X \subseteq S^\infty[d]$.

2. $\widehat{\mathcal{G}}(X)$ is the set of all $s \in [0, \infty)$ such that there is an $s$-supergale $d$ for which $X \subseteq S^\infty[d]$.

Note that $s' > s \in \mathcal{G}(X)$ implies that $s' \in \mathcal{G}(X)$.

It was shown in [25] that the following definition is equivalent to the classical definition of Hausdorff dimension in **C**.

**Definition.** The *Hausdorff dimension* of a set $X \subseteq \mathbf{C}$ is $\dim_H(X) = \inf \mathcal{G}(X)$.

The following trivial fact shows that we could equivalently use $\widehat{\mathcal{G}}(X)$ in place of $\mathcal{G}(X)$ in the above definition.

**Observation 3.4.** *For all $X \subseteq \mathbf{C}$, $\mathcal{G}(X) = \widehat{\mathcal{G}}(X)$.*

Martin-Löf's definition of randomness [26] was reformulated in terms of martingales by Schnorr [36] as follows.

**Terminology.** A $\nu$-$s$-supergale is *constructive* if it is lower semicomputable.

**Definition.** Let $\nu$ be a probability measure on **C**, and let $X \subseteq \mathbf{C}$.

1. $X$ has *constructive $\nu$-measure 0*, and we write $\nu_{\text{constr}}(X) = 0$, if there is a constructive $\nu$-martingale $d$ such that $X \subseteq S^\infty[d]$.

2. $X$ has *constructive $\nu$-measure 1*, and we write $\nu_{\text{constr}}(X) = 1$, if $\nu_{\text{constr}}(\mathbf{C} - X) = 0$.

**Definition.** If $\nu$ is a probability measure on **C**, then a sequence $R \in \mathbf{C}$ is *$\nu$-random*, and we write $R \in \text{RAND}_\nu$, if the singleton set $\{R\}$ does not have constructive $\nu$-measure 0 (i.e., there is no constructive $\nu$-martingale that succeeds on $R$).

It is well-known (and easy to see) that $\nu_{\text{constr}}(\text{RAND}_\nu) = 1$. The following known result shows that constructive supermartingales can equivalently be used in place of constructive martingales in defining randomness.

**Theorem 3.5.** (Schnorr [36, 37]) *For every computable probability measure $\nu$ on **C** and every constructive $\nu$-supermartingale $d$ there is a constructive $\nu$-martingale $d'$ such that $S^\infty[d] \subseteq S^\infty[d']$.*

If $\nu$ is $\mu$, the uniform probability measure on **C**, then we generally omit it from the above terminology. A sequence $R$ is thus *random*, and we write $R \in \text{RAND}$, if $\{R\}$ does not have constructive measure 0.

Optimal constructive supergales are as crucial to constructive dimension as optimal constructive supermartingales are to the theory of randomness. Recall the subprobability supermeasure **M** of Theorem 2.1.

**Notation.** For each $s \in [0, \infty)$ and $w \in \{0, 1\}^*$,
$$\mathbf{d}^{(s)}(w) = 2^{s|w|} \mathbf{M}(w).$$

**Theorem 3.6.** *For every computable real number $s \in [0, \infty)$, the function $\mathbf{d}^{(s)}$ is an optimal constructive $s$-supergale.*

*Proof.* Let $s \in [0, \infty)$ be computable. It is clear from its definition that $\mathbf{d}^{(s)}$ is a constructive $s$-supergale. To see that $\mathbf{d}^{(s)}$ has the desired optimality property, let $d$ be an arbitrary constructive $s$-supergale. Fix $0 < a \in \mathbb{Q}$ such that $ad(\lambda) \leq 1$, and define $\nu : \{0,1\}^* \to [0,1]$ by

$$\nu(w) = 2^{-s|w|}ad(w)$$

for all $w \in \{0,1\}^*$. By Observation 3.2, $\nu$ is a 0-supergale. Since $\nu(\lambda) = ad(\lambda) \leq 1$, it follows by Observation 3.1 that $\nu$ is a subprobability supermeasure on $\mathbf{C}$. Since $\nu$ is clearly constructive, it follows by the optimality of $\mathbf{M}$ that there exists $\alpha > 0$ such that for all $w \in \{0,1\}^*$, $\mathbf{M}(w) \geq \alpha\nu(w)$, whence

$$\mathbf{d}^{(s)}(w) = 2^{s|w|}\mathbf{M}(w) \geq 2^{s|w|}\alpha\nu(w) = \alpha ad(w).$$

Since $\alpha a > 0$ this shows that $\mathbf{d}^{(s)}$ is an optimal constructive $s$-supergale. □

We now constructivize the above definition of Hausdorff dimension and develop some fundamental properties of the resulting constructive dimension.

**Notation.** Let $X \subseteq \mathbf{C}$.

1. $\mathcal{G}_{\text{constr}}(X)$ is the set of all $s \in [0, \infty)$ such that there is a constructive $s$-gale $d$ for which $X \subseteq S^\infty[d]$.

2. $\widehat{\mathcal{G}}_{\text{constr}}(X)$ is the set of all $s \in [0, \infty)$ such that there is constructive $s$-supergale $d$ for which $X \subseteq S^\infty[d]$.

Note that if $s, s'$ are computable real numbers with $s' > s$, then $s \in \mathcal{G}_{\text{constr}}(X)$ implies $s' \in \mathcal{G}_{\text{constr}}(X)$, and $s \in \widehat{\mathcal{G}}_{\text{constr}}(X)$ implies $s' \in \widehat{\mathcal{G}}_{\text{constr}}(X)$.

We have seen that gales and supergales can be used interchangeably in defining classical Hausdorff dimension (Observation 3.4) and that constructive martingales and constructive supermartingales can be used interchangeably in defining randomness (Theorem 3.5). In contrast, at the time of this writing, we do not know whether constructive gales and constructive supergales can be used interchangeably in defining constructive dimension. We also do not know whether an analog of Theorem 3.6 holds for constructive $s$-gales when $s < 1$. We thus define constructive dimension in terms of constructive supergales.

**Definition.** The *constructive dimension* of a set $X \subseteq \mathbf{C}$ is $\text{cdim}(X) = \inf \widehat{\mathcal{G}}_{\text{constr}}(X)$.

The following observations are clear.

**Observations 3.7.** *1. For all $X \subseteq Y \subseteq \mathbf{C}$, $\text{cdim}(X) \leq \text{cdim}(Y)$.*

*2. For all $X \subseteq \mathbf{C}$, $\text{cdim}(X) \geq \dim_\text{H}(X)$.*

*3. $\text{cdim}(\mathbf{C}) = 1$.*

*4. For all $X \subseteq \mathbf{C}$, if $\text{cdim}(X) < 1$, then $\mu_{\text{constr}}(X) = 0$.*

# 4 Dimensions of Individual Sequences

The dimension of an individual sequence $S \in \mathbf{C}$ is simply the constructive dimension of the singleton set $\{S\}$.

**Definition.** The *dimension* of a sequence $S \in \mathbf{C}$ is

$$\dim(S) = \operatorname{cdim}(\{S\}).$$

The following theorem, which has no analog either in classical Hausdorff dimension or in the resource-bounded dimension developed in [25], says that the constructive dimension of a set of sequences is completely determined by the dimensions of the individual sequences in the set.

**Theorem 4.1.** *For all $X \subseteq \mathbf{C}$,*

$$\operatorname{cdim}(X) = \sup_{S \in X} \dim(S).$$

*Proof.* Let $X \subseteq \mathbf{C}$, and let $s^* = \sup_{s \in X} \dim(S)$. It is clear by part 1 of Observation 3.7 that $\operatorname{cdim}(X) \geq s^*$. To see that $\operatorname{cdim}(X) \leq s^*$, let $s$ be a rational number such that $s > s^*$. It suffices to show that $\operatorname{cdim}(X) \leq s$.

Since $s > s^*$, for each $S \in X$ there is a constructive $s$-supergale $d_S$ that succeeds on $S$. By Theorem 3.6, then, we have $S \in S^{\infty}[d_S] \subseteq S^{\infty}[\mathbf{d}^{(s)}]$ for all $S \in X$, whence $X \subseteq S^{\infty}[\mathbf{d}^{(s)}]$. Since $\mathbf{d}^{(s)}$ is a constructive $s$-supergale, this shows that $\operatorname{cdim}(X) \leq s$. □

Hitchcock [14] has recently proven a correspondence principle for constructive dimension. This principle says that for any set $X \subseteq \mathbf{C}$ that is a union of $\Pi_1^0$ sets (a condition that is certainly satisfied if $X$ is $\Sigma_2^0$), the constructive dimension of $X$ is precisely its classical Hausdorff dimension. He also noted that this principle, together with Theorem 4.1, implies that the *classical* Hausdorff dimension of every set $X \subseteq \mathbf{C}$ that is a union of $\Pi_1^0$ sets has the *pointwise characterization*

$$\dim_{\mathrm{H}}(X) = \sup_{S \in X} \dim(S).$$

Theorem 4.1 immediately implies that constructive dimension has the following *countable stability* property, which is also a property of classical Hausdorff dimension.

**Corollary 4.2.** *For all $X_0, X_1, X_2, \ldots \subseteq \mathbf{C}$,*

$$\operatorname{cdim}\left(\bigcup_{k=0}^{\infty} X_k\right) = \sup_{k \in \mathbb{N}} \operatorname{cdim}(X_k).$$

Our next objective is to prove a dimension reduction theorem that enables us to exhibit sequences of arbitrary $\Delta_2^0$-computable dimensions in [0,1].

Define an *approximator* of a real number $\alpha \in [0,1]$ to be an ordered pair $(a,b)$ of computable functions $a, b : \mathbb{N} \to \mathbb{Z}^+$ with the following properties.

(i) For all $n \in \mathbb{N}$, $a(n) \leq b(n)$.

(ii) $\lim_{n \to \infty} \frac{a(n)}{b(n)} = \alpha$.

It is well known and easy to see that a real number $\alpha \in [0,1]$ has an approximator if and only if it is $\Delta_2^0$-computable. Moreover, every $\Delta_2^0$-computable real number has an approximator $(a,b)$ that is *nice* in the sense that if we let $\tilde{b}(k) = \sum_{n=0}^{k-1} b(n)$, then $b(k) = o(\tilde{b}(k))$ as $k \to \infty$.

Given an approximator $(a,b)$ of a $\Delta_2^0$-computable real number $\alpha \in [0,1]$, we define the $(a,b)$-*dilution function*

$$g_{(a,b)} : \mathbf{C} \to \mathbf{C}$$

as follows. Given $S \in \mathbf{C}$, if we write

$$S = w_0 w_1 w_2 \ldots,$$

where $|w_n| = a(n)$ for each $n \in \mathbb{N}$, then

$$g_{(a,b)}(S) = w_0 0^{b(0)-a(0)} w_1 0^{b(1)-a(1)} \ldots.$$

Note that $g_{(a,b)}(S) \equiv_1 S$ for all $S \in \mathbf{C}$.

**Theorem 4.3.** *Let $\alpha \in [0,1]$ be $\Delta_2^0$-computable, and let $(a,b)$ be a nice approximator of $\alpha$. Then for all $S \in \mathbf{C}$,*

$$\dim(g_{(a,b)}(S)) = \alpha \cdot \dim(S).$$

*Proof.* We first introduce some notation that will simplify the proof. Let $(a,b)$ be a nice approximator of $\alpha$. For each $k \in \mathbb{N}$, let

$$\widetilde{a}(k) \sum_{n=0}^{k-1} a(n), \quad \widetilde{b}(k) = \sum_{n=0}^{k-1} b(n),$$

and note that

$$\lim_{k \to \infty} \frac{\widetilde{a}(k)}{\widetilde{b}(k)} = \alpha.$$

In addition to the dilution function $g_{(a,b)} : \mathbf{C} \to \mathbf{C}$, we use the function $g : \{0,1\}^* \to \{0,1\}^*$ defined recursively as follows. First, $g(\lambda) = \lambda$. Next, if $w = w'u$, where $|w'| = \widetilde{a}(k)$ and $0 < |u| < a(k)$, then $g(w) = g(w')u$. Finally, if $w = w'u$, where $|w'| = \widetilde{a}(k)$ and $|u| = a(k)$, then $g(w) = g(w')u0^{b(k)-a(k)}$. Note that for all $S \in \mathbf{C}$, $g_{(a,b)}(S)$ is the unique $T \in \mathbf{C}$ such that $g(w) \sqsubseteq T$ for all $w \sqsubseteq S$. Note also that the function $g$ is one-to-one, so that the string $g^{-1}(y)$ is well defined for each $y \in \text{range}(g)$.

Now fix $S \in \mathbf{C}$ and let $\beta = \dim(S)$. Our objective is to show that $\dim(g_{(a,b)}(S)) = \alpha\beta$.

To see that $\dim(g_{(a,b)}(S)) \leq \alpha\beta$, let $s > \beta$ and $t > \alpha$ be such that $2^s$ and $2^t$ are rational. It suffices to show that $\dim(g(S)) \leq st$.

Since $s > \beta$, there is a constructive $s$-supergale $d_S$ that succeeds on $S$. Define a function $d : \{0,1\}^* \to [0,\infty)$ as follows. Let $y \in \{0,1\}^*$. If there does not exist $T \in \mathbf{C}$ such that $y \sqsubseteq g_{(a,b)}(T)$, then $d(y) = 0$. Otherwise, let $w$ be the shortest string such that $y \sqsubseteq g(w)$. Then

$$d(y) = 2^{st|y|-s|w|} d_S(w).$$

It is routine to check that $d$ is an $st$-supergale, and it is clear that $d$ is constructive. Also, for each $w \in \{0,1\}^*$,

$$d(g(w)) = 2^{st|g(w)|-s|w|} d_S(w). \tag{4.1}$$

Let $\epsilon = \frac{t-\alpha}{4}$ and fix $k_0 \in \mathbb{N}$ such that for all $k \geq k_0$,

$$\frac{\widetilde{a}(k)}{\widetilde{b}(k)} \leq t - 2\epsilon \text{ and } b(k) \leq \epsilon \widetilde{b}(k).$$

(Such $k_0$ exists because $\frac{\widetilde{a}(k)}{\widetilde{b}(k)}$ converges to $\alpha$ and the approximator $(a,b)$ is nice.) For all $w \in \{0,1\}^*$, if we choose $k$ and $r$ such that $|w| = \widetilde{a}(k) + r$ and $0 \leq r < a(k)$, and if $k \geq k_0$, then we have

$$\begin{aligned}
st|g(w)| - s|w| &= st(\widetilde{b}(k) + r) - s(\widetilde{a}(k) + r) \\
&= st\widetilde{b}(k) - s\widetilde{a}(k) - s(1-t)r \\
&\geq st\widetilde{b}(k) - s(t-2\epsilon)\widetilde{b}(k) - s\epsilon\widetilde{b}(k) \\
&= \epsilon\widetilde{b}(k).
\end{aligned}$$

Since $\epsilon > 0$ and $S \in S^\infty[d_S]$, it follows by (4.1) that $g_{(a,b)}(S) \in S^\infty[d]$. Since $d$ is a constructive $st$-supergale, this establishes that $\dim(g_{(a,b)}(S)) \leq st$, concluding the proof that $\dim(g_{(a,b)}(S)) \leq \alpha\beta$.

To see that $\dim(g_{(a,b)}(S)) \geq \alpha\beta$, let $s < \alpha\beta$ be such that $2^s$ is rational, and let $d$ be a constructive $s$-supergale. It suffices to show that $g_{(a,b)}(S) \notin S^\infty[d]$.

Define a function $d' : \{0,1\}^* \to [0, \infty)$ by

$$d'(w) = 2^{s|w| - s|g(w)|} d(g(w)).$$

Using Lemma 3.3, it is easy to check that $d'$ is a constructive $s$-supergale. Since $s < \alpha\beta$, we can choose $t < \beta$ such that $s < \alpha t$ and $2^t$ is rational. The function $d'' : \{0,1\}^* \to [0, \infty)$ defined by

$$d''(w) = 2^{(t-s)|w|} d'(w)$$

is then a constructive $t$-supergale by Observation 3.2. Since $t < \beta = \dim(S)$, it follows that there is a constant $c \in \mathbb{N}$ such that for all $w \sqsubseteq S$, $d''(w) \leq 2^c$.

Let $\epsilon = \frac{t}{2}(\alpha - \frac{s}{t})$, noting that this is positive because $s < \alpha t$. Fix $k_0 \in \mathbb{N}$ such that for all $k \geq k_0$,

$$\frac{\widetilde{a}(k)}{\widetilde{b}(k)} \geq \frac{s + \epsilon}{t} \text{ and } b(k) \leq \epsilon \widetilde{b}(k).$$

(Such $k_0$ exists because $\frac{\widetilde{a}(k)}{\widetilde{b}(k)}$ converges to $\alpha$, $\frac{s+\epsilon}{t} < \alpha$, and the approximator $(a, b)$ is nice.) Every $y \sqsubseteq g_{(a,b)}(S)$ can be written in the form $y = g(w)u$, where $w \sqsubseteq S$, $|w| = \widetilde{a}(k)$, $|g(w)| = \widetilde{b}(k)$, and $|u| < b(k)$. For such $y$ we have

$$\begin{aligned}
d(y) &\leq 2^{s|u|} d(g(w)) \leq 2^{b(k)} d(g(w)) \\
&= 2^{b(k) - s|w| + s|g(w)|} d'(w) \\
&= 2^{b(k) + s|g(w)| - t|w|} d''(w) \\
&= 2^{b(k) + s\widetilde{b}(k) - t\widetilde{a}(k)} d''(w) \\
&\leq 2^{b(k) + s\widetilde{b}(k) - t\widetilde{a}(k) + c}.
\end{aligned}$$

If $|y| \geq \widetilde{b}(k_0)$, so that $k \geq k_0$, then we have

$$b(k) + s\widetilde{b}(k) - t\widetilde{a}(k) \leq \epsilon \widetilde{b}k + s\widetilde{b}(k) - (s+\epsilon)\widetilde{b}(k) = 0,$$

whence

$$d(y) \leq 2^c.$$

Since this holds for all sufficiently long prefixes $y \sqsubseteq g_{(a,b)}(S)$, it follows that $g_{(a,b)}(S) \notin S^\infty[d]$, concluding the proof that $\dim(g_{(a,b)}(S)) \geq \alpha\beta$. $\square$

**Notation.** For each $\alpha \in [0, 1]$, let

$$\begin{aligned}
\text{DIM}_\alpha &= \{S \in \mathbf{C} \mid \dim(S) = \alpha\}, \\
\text{DIM}_{\leq \alpha} &= \{S \in \mathbf{C} \mid \dim(S) \leq \alpha\}, \\
\text{DIM}_{<\alpha} &= \{S \in \mathbf{C} \mid \dim(S) < \alpha\}.
\end{aligned}$$

**Observation 4.4.** $\text{RAND} \subseteq \text{DIM}_1$.

*Proof.* This follows immediately from part 4 of Observation 3.7. $\square$

An important result in the theory of random sequences is the existence of random sequences in $\Delta_2^0$. We now use this fact and Theorem 4.3 to show that there are $\Delta_2^0$ sequences of every $\Delta_2^0$-computable dimension in [0,1].

**Theorem 4.5.** *For every $\Delta_2^0$-computable real number $\alpha \in [0,1]$, $\text{DIM}_\alpha \cap \Delta_2^0 \neq \emptyset$, i.e., there is a $\Delta_2^0$ sequence $S$ such that $\dim(S) = \alpha$.*

*Proof.* Let $\alpha \in [0,1]$ be $\Delta_2^0$-computable. It is well known and easy to see that $\alpha$ has an approximator (indeed, this characterizes $\Delta_2^0$-computability), and it is routine to transform an approximator of $\alpha$ into a nice approximator $(a,b)$ of $\alpha$. It is well known (see [49, 50] or [22]) that there is a sequence $R \in \text{RAND} \cap \Delta_2^0$. Let $S = g_{(a,b)}(R)$. Then Theorem 4.3 and Observation 4.4 tell us that

$$\dim(S) = \alpha \dim(R) = \alpha.$$

□

Three remarks on the proof of Theorem 4.5 should be made here. First, the proof that $\text{RAND} \cap \Delta_2^0 \neq \emptyset$ using Kreisel's Basis Lemma [19, 49, 50, 30] and the fact that RAND is a $\Sigma_2^0$ set cannot directly be adapted to proving that $\text{DIM}_\alpha \cap \Delta_2^0 \neq \emptyset$ because Terwijn [48] has shown that $\text{DIM}_\alpha$ is not a $\Sigma_2^0$ set. Second, Mayordomo [28] has recently generalized Chaitin's $\Omega$ construction [6] to give an alternative construction of sequences in $\text{DIM}_\alpha \cap \Delta_2^0$. Third, our proof of Theorem 4.5 via Theorem 4.3 yields even more, namely, that if $\alpha, \beta \in [0,1]$ are $\Delta_2^0$-computable with $\alpha \geq \beta$, then every sequence in $\text{DIM}_\alpha$ is 1-equivalent to some sequence in $\text{DIM}_\beta$.

The following theorem shows that Theorem 4.5 cannot be improved to $\Sigma_1^0$ or $\Pi_1^0$ sequences.

**Theorem 4.6.** $\Sigma_1^0 \cup \Pi_1^0 \subseteq \text{DIM}_0$.

*Proof.* Let $S \in \Sigma_1^0$. By symmetry, it suffices to show that $\dim(S) = 0$. For this, let $0 < s \in \mathbb{Q}$. It suffices to show that $\dim(S) \leq s$.

By standard techniques [31, 42], let $S_0, S_1, \ldots$ be a sequence of elements of **C** with the following properties.

(i) For each $t$, $S_t$ contains only finitely many 1's.

(ii) For each $t$ and $n$, $S_t[n] \leq S_{t+1}[n]$.

(iii) For each $n$, $S[n] = \lim_{t \to \infty} S_t[n]$.

(iv) The set $\{(t,n)|S_t[n] = 1\}$ is computably enumerable.

That is, $S_t$ is the "$t^{\text{th}}$ finite approximation of $S$."

Define a function $d : \{0,1\}^* \to [0,\infty)$ as follows. First, $d(\lambda) = 1$. Next, assume that $d(w)$ has been defined, where $|w| = \binom{n}{2}$ for some integer $n \geq 1$. For each $u \in \{0,1\}^n$, define

$$d(wu) = \begin{cases} \frac{2^{sn}d(w)}{n+1} & \text{if } (\exists t) S_t[\binom{n}{2}..\binom{n+1}{2} - 1] = u \\ 0 & \text{otherwise,} \end{cases}$$

noting that $|wu| = \binom{n}{2} + n = \binom{n+1}{2}$. For each $u$ such that $0 < |u| < n$ define

$$d(wu) = 2^{-s(n-|u|)} \sum_{|v|=n-|u|} d(wuv).$$

Since there are at most $n+1$ strings $u \in \{0,1\}^n$ for which $d(wu) > 0$, it is clear that $d$ is an $s$-supergale. It is also clear that $d$ is constructive and that $d$ succeeds on $S$, whence $\dim(S) \leq s$. □

The rest of this section concerns the constructive dimensions of the dimension classes $\text{DIM}_{\leq \alpha}$ and $\text{DIM}_{<\alpha}$. We first note that for every $\alpha \in [0,1]$, $\text{DIM}_{\leq \alpha}$ is the largest set of constructive dimension $\alpha$.

**Theorem 4.7.** *For every $\alpha \in [0,1]$, the set $\text{DIM}_{\leq \alpha}$ has the following two properties.*

1. $\text{cdim}(\text{DIM}_{\leq \alpha}) = \alpha$.

2. *For all $X \subseteq \mathbf{C}$, if $\text{cdim}(X) \leq \alpha$, then $X \subseteq \text{DIM}_{\leq \alpha}$.*

*Proof.* Part 1 follows immediately from Theorem 4.1, Theorem 4.5, and the fact that the $\Delta_2^0$-computable reals are dense in $\mathbb{R}$. Part 2 follows immediately from part 1 of Observation 3.7. □

Part 1 of Theorem 4.7 has the following immediate consequence.

**Corollary 4.8.** *For every $\alpha \in [0,1]$,*

$$\text{cdim}(\text{DIM}_{<\alpha}) = \alpha.$$

We show in section 6 below that

$$\text{cdim}(\text{DIM}_\alpha) = \alpha$$

for all reals $\alpha \in [0,1]$.

## 5 Dimensions of Individual Strings

In the preceding two sections we have constructivized classical Hausdorff dimension and thereby defined the dimensions of individual infinite binary sequences. We now push this one step further by constructivizing *and discretizing* classical Hausdorff dimension in order to define the dimensions of individual finite binary strings.

Recall that the dimension of a sequence $S$ is the infimum of all $s \geq 0$ for which there exists a constructive $s$-supergale $d$ such that the values of $d(S[0..n-1])$ are unbounded as $n \to \infty$. To define the dimensions of finite strings, we modify this definition in three ways.

I. We replace supergales by termgales, which are supergale-like constructs with special requirements for handling the terminations of strings.

II. We replace "unbounded as $n \to \infty$" by a finite threshold.

III. We make the definition universal by using an optimal constructive termgale.

We now carry out this development.

Supergales are well suited to defining the dimensions of infinite sequences, but an adequate definition of the dimensions of finite strings must also involve betting on the point at which a given string terminates. We use the termination symbol □ to mark the end of a binary string. We work in the set

$$\mathcal{T} = \{0,1\}^* \cup \{0,1\}^*\square,$$

consisting of all *terminated binary strings* (elements of $\{0,1\}^*\square$) and prefixes thereof. The following definition is the main idea of this section.

**Definition.** For $s \in [0, \infty)$, an *s-termgale* is a function
$$d : \mathcal{T} \to [0, \infty)$$
such that $d(\lambda) \leq 1$ and for all $w \in \{0,1\}^*$,
$$d(w) \geq 2^{-s}[d(w0) + d(w1) + d(w\square)]. \tag{5.1}$$

An $s$-termgale $d$ is a strategy for betting on the successive bits of a binary string and also on the point at which the string terminates. We require the initial capital $d(\lambda)$ to be at most 1. When $d$ is used to bet on a string $w$, the final capital is $d(w\square)$.

The payoff condition (5.1) may at first glance seem suspicious. In the case $s = 1$, this says that
$$d(w) \geq \frac{d(w0) + d(w1) + d(w\square)}{2} \tag{5.2}$$
for all $w \in \{0,1\}^*$. If each of 0, 1, and $\square$ is equally likely to occur, independently of all prior bits, then (5.2) implies that the conditional expected capital after a bet, given that $w$ has occurred before the bet, is
$$\frac{d(w0) + d(w1) + d(w\square)}{3} = \frac{2}{3}d(w),$$
whence the payoffs are much less than fair, even if equality holds in (5.2). However, the assumption that 0, 1, and $\square$ are equally likely to occur is not reasonable because it forces strings to be very short with overwhelming probability. (In fact, this assumption implies that the average string is only two bits long. In contrast, the average length of a string with respect to the optimal constructive subprobability measure **m** is infinite.) Since we want our theory to apply to long strings, the termination symbol $\square$ should be regarded as having a vanishingly small probability.

The 1-termgale payoff condition (5.2) is exactly the supermartingale (i.e., 1-supergale) payoff condition with the additional requirement that the 1-termgale must without compensation divert some of its capital to bet on $\square$, i.e., the possibility that there is no next bit. Since $\square$ can only occur once, the overall impact of this requirement is modest. However, the impact is real, and we shall see that it is exactly what is needed.

**Example 5.1.** Define $d : \mathcal{T} \to [0, \infty)$ by the recursion
$$\begin{aligned} d(\lambda) &= 1, \\ d(w0) &= \tfrac{3}{2}d(w), \\ d(w1) &= d(w\square) = \tfrac{1}{4}d(w). \end{aligned}$$

It is clear that $d$ is a 1-termgale. If $w$ is a binary string of length $n$ with $n_0$ 0's and $n_1$ 1's, then
$$\begin{aligned} d(w\square) &= (\tfrac{3}{2})^{n_0}(\tfrac{1}{4})^{n_1+1} \\ &= 2^{n_0(1+\log 3) - 2(n+1)}. \end{aligned}$$

In particular, if $n_0$ is significantly larger than $\frac{2}{1+\log 3}(n+1) \approx 0.7737(n+1)$, then $d(w\square)$ is significantly greater than $d(\lambda)$ even though $d$ loses three-fourths of its capital when the $\square$ appears.

The following analog of Observation 3.2 is obvious but useful.

**Observation 5.2.** *Let $d, d' : \mathcal{T} \to [0, \infty)$ and $s, s' \in [0, \infty)$. If*
$$2^{-s|x|}d(x) = 2^{-s'|x|}d'(x)$$
*for all $x \in \mathcal{T}$, then $d$ is an $s$-termgale if and only if $d'$ is an $s'$-termgale.*

In particular, if $d$ is a 0-termgale and $s \in [0, \infty)$, then the function $d'$ defined by
$$d'(x) = 2^{s|x|}d(x)$$
for all $x \in \mathcal{T}$ is an $s$-termgale, and every $s$-termgale can be obtained from a 0-termgale in this way.

**Lemma 5.3.** *If $s \in [0, \infty)$ and $d$ is an $s$-termgale, then for all $u \in \{0,1\}^*$,*
$$\sum_{w \in \{0,1\}^*} 2^{-s|w|}d(uw\square) \leq 2^s d(u). \tag{5.3}$$

*Proof.* For the first part of the proof, assume that $d$ is a 0-termgale, and let $u \in \{0,1\}^*$. We begin by using induction on $m$ to show that
$$\sum_{w \in \{0,1\}^{<m}} d(uw\square) + \sum_{w \in \{0,1\}^m} d(uw) \leq d(u) \tag{5.4}$$
for all $m \in \mathbb{N}$. For $m = 0$, this is trivial. Assume that it holds for $m$. Then
$$\sum_{w \in \{0,1\}^{<m+1}} d(uw\square) + \sum_{w \in \{0,1\}^{m+1}} d(uw)$$
$$= \sum_{w \in \{0,1\}^{<m}} d(uw\square) + \sum_{w \in \{0,1\}^m} d(uw\square) + \sum_{w \in \{0,1\}^m} [d(uw0) + d(uw1)]$$
$$\leq \sum_{w \in \{0,1\}^{<m}} d(uw\square) + \sum_{w \in \{0,1\}^m} d(uw)$$
$$\leq d(u)$$
by the induction hypothesis. This confirms that (5.4) holds for all $m \in \mathbb{N}$. It follows immediately that
$$\sum_{w \in \{0,1\}^{\leq m}} d(uw\square) \leq d(u)$$
for all $m \in \mathbb{N}$, whence
$$\sum_{w \in \{0,1\}^*} d(uw\square) \leq d(u).$$
This is the case $s = 0$ of (5.3).

Now assume that $d$ is an $s$-termgale, where $s \in [0, \infty)$ is arbitrary. Define $d' : \mathcal{T} \to [0, \infty)$ by $d'(x) = 2^{-s|x|}d(x)$ for all $x \in \mathcal{T}$. Then $d'$ is a 0-termgale by Observation 5.2, so the first part of this proof tells us that for all $u \in \{0,1\}^*$,
$$\sum_{w \in \{0,1\}^*} 2^{-s|w|}d(uw\square) = 2^{s|u\square|} \sum_{w \in \{0,1\}^*} d'(uw\square)$$
$$\leq 2^{s|u\square|}d'(u)$$
$$= 2^s d(u).$$
$\square$

To define optimal termgales we need uniformity in the parameter $s$.

**Definition.** 1. A *termgale* is a family $d = \{d^{(s)} | s \in [0, \infty)\}$ such that each $d^{(s)}$ is an $s$-termgale and for all $s, s' \in [0, \infty)$ and $x \in \mathcal{T}$,
$$2^{-s|x|}d^{(s)}(x) = 2^{-s'|x|}d^{(s')}(x).$$

2. A termgale $d$ is *constructive* if $d^{(0)}$ is constructive.

**Definition.** A constructive termgale $\tilde{d}$ is *optimal* if for every constructive termgale $d$ there is a constant $\alpha > 0$ such that for all $s \in [0, \infty)$ and $w \in \{0,1\}^*$, $\tilde{d}^{(s)}(w\square) \geq \alpha d^{(s)}(w\square)$.

**Definition.** The *termgale induced by* a subprobability measure $p$ on $\{0,1\}^*$ is the family $d[p] = \{d[p]^{(s)} | s \in [0, \infty)\}$, where each $d[p]^{(s)} : \mathcal{T} \to [0, \infty)$ is defined by

$$d[p]^{(s)}(x) = 2^{s|x|} \sum_{\substack{w \in \{0,1\}^* \\ x \sqsubseteq w\square}} p(w)$$

for all $x \in \mathcal{T}$.

**Theorem 5.4.** *If $\mathbf{p}$ is an optimal constructive subprobability measure on $\{0,1\}^*$, then $d[\mathbf{p}]$ is an optimal constructive termgale.*

*Proof.* Assume the hypothesis. It is clear that $d[\mathbf{p}]$ is a constructive termgale. To see that $d[\mathbf{p}]$ is optimal, let $d = \{d^{(s)} | s \in [0, \infty)\}$ be an arbitrary constructive termgale. Define $p : \{0,1\}^* \to [0, \infty)$ by $p(w) = d^{(0)}(w\square)$ for all $w \in \{0,1\}^*$. By Lemma 5.3 (with $u = \lambda$), $p$ is a subprobability measure on $\{0,1\}^*$, and $p$ is constructive because $d$ is constructive. It follows by the optimality of $\mathbf{p}$ that there exists $\alpha > 0$ such that $\mathbf{p}(w) \geq \alpha p(w)$ for all $w \in \{0,1\}^*$. Then for all $s \in [0, \infty)$ and $w \in \{0,1\}^*$,

$$\begin{aligned} d[\mathbf{p}]^{(s)}(w\square) &= 2^{s|w\square|}\mathbf{p}(w) \\ &\geq 2^{s|w\square|}\alpha p(w) \\ &= 2^{s|w\square|}\alpha d^{(0)}(w\square) \\ &= \alpha d^{(s)}(w\square), \end{aligned}$$

so $d[\mathbf{p}]$ is optimal. $\square$

**Corollary 5.5.** *There exists an optimal constructive termgale.*

*Proof.* This follows immediately from Theorems 5.4 and 2.1. $\square$

We can now implement the ideas I, II, and III described at the beginning of this section.

**Definition.** If $d$ is a termgale, $l \in \mathbb{Z}^+$, and $w \in \{0,1\}^*$, then the *dimension* of $w$ relative to $d$ at significance level $l$ is
$$\dim_d^l(w) = \inf\{s \in [0, \infty) \mid d^{(s)}(w\square) > l\}.$$
We write $\dim_d(w)$ for $\dim_d^1(w)$.

**Theorem 5.6.** *If $\tilde{d}$ is an optimal constructive termgale, then for every constructive termgale $d$ and every $l \in \mathbb{Z}^+$, there is a constant $\gamma \in [0, \infty)$ such that for all $w \in \{0,1\}^*$,*

$$\dim_{\tilde{d}}^l(w) \leq \dim_d(w) + \frac{\gamma}{1 + |w|}. \tag{5.5}$$

*Proof.* Let $\tilde{d}$ be an optimal constructive termgale, let $d$ be an arbitrary constructive termgale, and let $l \in \mathbb{Z}^+$. By the optimality of $\tilde{d}$, there is a constant $\alpha \in (0, 1]$ such that for all $s \in [0, \infty)$ and $w \in \{0,1\}^*$, $\tilde{d}^{(s)}(w\square) \geq \alpha d^{(s)}(w\square)$. Let $\gamma = \log l - \log \alpha$, and note that $\gamma \in [0, \infty)$. Let $w \in \{0,1\}^*$ be arbitrary. To see that (5.5) holds, let $s > \dim_d(w) + \frac{\gamma}{1+|w|}$. It suffices to show that $\tilde{d}^{(s)}(w\square) > l$.

Let $s_1 = s - \frac{\gamma}{1+|w|}$. Then $s_1 > \dim_d(w)$, so

$$\begin{aligned} \tilde{d}^{(s)}(w\square) &\geq \alpha d^{(s)}(w\square) \\ &= \alpha 2^{(s-s_1)|w\square|} d^{(s_1)}(w\square) \\ &= \alpha 2^{\gamma} d^{(s_1)}(w\square) \\ &> \alpha 2^{\gamma} \\ &= l. \end{aligned}$$

□

**Corollary 5.7.** *If $\tilde{d}_1$ and $\tilde{d}_2$ are optimal constructive termgales and $l_1, l_2 \in \mathbb{Z}^+$, then there is a constant $\alpha \in [0, \infty)$ such that for all $w \in \{0,1\}^*$,*

$$\left| \dim_{\tilde{d}_1}^{l_1}(w) - \dim_{\tilde{d}_2}^{l_2}(w) \right| \leq \frac{\alpha}{1+|w|}.$$

Corollary 5.7 says that if we base our definition of dimension on an optimal constructive termgale $\tilde{d}$, then both the particular choice of $\tilde{d}$ and the choice of a significance level $l$ have negligible impact on the dimension $\dim_{\tilde{d}}^l(w)$. We thus fix an optimal constructive termgale $\mathbf{d}_\square$ and define the dimensions of finite strings as follows.

**Definition.** The *dimension* of a string $w \in \{0,1\}^*$ is

$$\dim(w) = \dim_{\mathbf{d}_\square}(w).$$

We have seen that the dimension of a sequence is at most 1. In contrast, we will see in section 6 that the dimension of a string may exceed 1. However, regardless of our choice of $\mathbf{d}_\square$, there is an upper bound on the dimension of strings.

**Lemma 5.8.** *There is a constant $c \in \mathbb{N}$ such that for all $w \in \{0,1\}^*$, $\dim(w) \leq c$.*

*Proof.* For each $s \in [0, \infty)$, define $d^{(s)} : \mathcal{T} \to [0, \infty)$ by

$$d^{(s)}(x) = \begin{cases} 2^{(s-2)|x|} & \text{if } x \in \{0,1\}^* \\ 2^{(s-2)|x|+1} & \text{if } x \in \{0,1\}^*\square, \end{cases}$$

and let $d = \{d^{(s)} | s \in [0, \infty)\}$. It is easy to see that $d$ is a constructive termgale and $d^{(2)}(w\square) = 2$ for all $w \in \{0,1\}^*$, whence $\dim_d(w) \leq 2$ for all $w \in \{0,1\}^*$. It follows by Theorem 5.6 that there is a constant $\gamma \in [0, \infty)$ such that for all $w \in \{0,1\}^*$,

$$\dim(w) \leq 2 + \frac{\gamma}{1+|w|} \leq 2 + \gamma.$$

Thus the present lemma holds with $c = 2 + \lceil \gamma \rceil$. □

We conclude this section by characterizing the dimension of a sequence in terms of the dimensions of its finite prefixes.

**Theorem 5.9.** *For all $S \in \mathbf{C}$,*

$$\dim(S) = \liminf_{n \to \infty} \dim(S[0..n-1]).$$

*Proof.* Let $S \in \mathbf{C}$. To see that $\dim(S) \leq \liminf_{n \to \infty} \dim(S[0..n-1])$, let $s$ and $s'$ be rational numbers such that $s' > s > \liminf_{n \to \infty} \dim(S[0..n-1])$. It suffices to show that $\dim(S) \leq s'$. By our choice of $s$, there is an infinite set $J \subseteq \mathbb{N}$ such that for all $n \in J$, $\dim(S[0..n-1]) < s$, whence $\mathbf{d}_\square^{(s)}(S[0..n-1]\square) > 1$. Define $d' : \{0,1\}^* \to [0, \infty)$ by $d'(w) = \mathbf{d}_\square^{(s')}(w)$ for all $w \in \{0,1\}^*$. Then $d'$ is a constructive $s'$-supergale and for all $n \in J$,

$$\begin{aligned}
d'(S[0..n-1]) &= \mathbf{d}_\square^{(s')}(S[0..n-1]) \\
&= 2^{(s'-s)n} \mathbf{d}_\square^{(s)}(S[0..n-1]) \\
&\geq 2^{(s'-s)n} 2^{-s} \mathbf{d}_\square^{(s)}(S[0..n-1]\square) \\
&> 2^{(s'-s)n-s}.
\end{aligned}$$

Since $J$ is infinite, this implies that $S \in S^\infty[d']$, whence $\dim(S) \leq s'$.

To see that $\dim(S) \geq \liminf_{n \to \infty} \dim(S[0..n-1])$, let $s'$ and $s''$ be rational numbers such that $s' > s'' > \dim(S)$. It suffices to show that there exist infinitely many $n \in \mathbb{N}$ for which $\dim(S[0..n-1]) \leq s'$. Since $s'' > \dim(S)$, there is a constructive $s''$-supergale $d$ such that $S \in S^\infty[d]$. Define $d' : \mathcal{T} \to [0, \infty)$ by

$$d'(x) = \begin{cases} d(x) & \text{if } x \in \{0,1\}^* \\ (2^{s'} - 2^{s''})d(w) & \text{if } x = w\square \in \{0,1\}^*\square. \end{cases}$$

Then $d'$ is a constructive $s'$-termgale, so if for each $s \in [0, \infty)$ we define $\tilde{d}^{(s)} : \mathcal{T} \to [0, \infty)$ by $\tilde{d}^{(s)}(x) = 2^{(s-s')|x|} d'(x)$, then the family $\tilde{d} = \{\tilde{d}^{(s)} | s \in [0, \infty)\}$ is a constructive termgale. It follows by the optimality of $\mathbf{d}_\square$ that there is a constant $\alpha > 0$ such that for all $s \in [0, \infty)$ and $w \in \{0,1\}^*$, $\mathbf{d}_\square^{(s)}(w\square) > \alpha \tilde{d}^{(s)}(w\square)$. Since $S \in S^\infty[d]$, there are infinitely many $n \in \mathbb{N}$ such that $\alpha(2^{s'} - 2^{s''})d(S[0..n-1]) > 1$. For all such $n$ we have

$$\begin{aligned}
\mathbf{d}_\square^{(s')}(S[0..n-1]\square) &\geq \alpha \tilde{d}^{(s')}(S[0..n-1]\square) \\
&= \alpha d'(S[0..n-1]\square) \\
&= \alpha(2^{s'} - 2^{s''})d(S[0..n-1]) \\
&> 1,
\end{aligned}$$

whence $\dim(S[0..n-1]) \leq s'$. $\square$

## 6 Dimension and Kolmogorov Complexity

In this section we show that the Kolmogorov complexity of a string is (up to an additive constant) the product of its length and its dimension. We use this to derive a new proof of a recent characterization of the dimension of a sequence in terms of the Kolmogorov complexities of its prefixes. This latter result is used to establish the existence of sequences of all dimensions in [0,1]. We also review some previous work on martingales, supermartingales, Kolmogorov complexity, and Hausdorff dimension.

**Theorem 6.1.** *There is a constant $c \in \mathbb{N}$ such that for all $w \in \{0,1\}^*$,*

$$\Big|K(w) - |w|\dim(w)\Big| \leq c.$$

*Proof.* Let **m** be the optimal subprobability measure on $\{0,1\}^*$ defined in section 2. The key fact is that for all $w \in \{0,1\}^*$ and $s \in [0, \infty)$,

$$d[\mathbf{m}]^{(s)}(w\square) > 1 \iff 2^{s|w\square|}\mathbf{m}(w) > 1$$
$$\iff s > \frac{1}{1+|w|} \log \frac{1}{\mathbf{m}(w)},$$

so

$$\dim_{d[\mathbf{m}]}(w) = \frac{1}{1+|w|} \log \frac{1}{\mathbf{m}(w)}.$$

This implies that

$$\log \frac{1}{\mathbf{m}(w)} = (1+|w|)\dim_{d[\mathbf{m}]}(w). \tag{6.1}$$

To complete the proof, fix constants $c_0, c_1, c_2 \in \mathbb{N}$ such that for all $w \in \{0,1\}^*$,

$$\left|K(w) - \log \frac{1}{\mathbf{m}(w)}\right| \leq c_0, \tag{6.2}$$

$$|\dim_{d[\mathbf{m}]}(w) - \dim(w)| \leq \frac{c_1}{1+|w|}, \tag{6.3}$$

and

$$\dim(w) \leq c_2. \tag{6.4}$$

(The constants $c_0$ and $c_2$ exist by Theorem 2.2 and Lemma 5.8, respectively. The constant $c_1$ exists by Theorem 5.4 and Corollary 5.7.) Let $c = c_0 + c_1 + c_2$. Then for all $w \in \{0,1\}^*$, (6.1) and (6.3) tell us that

$$\left|\log \frac{1}{\mathbf{m}(w)} - (1+|w|)\dim(w)\right| \leq c_1, \tag{6.5}$$

and (6.4) tells us that

$$|(1+|w|)\dim(w) - |w|\dim(w)| \leq c_2. \tag{6.6}$$

By (6.2), (6.5), (6.6), and the triangle inequality, we have

$$|K(w) - |w|\dim(w)| \leq c$$

for all $w \in \{0,1\}^*$. □

In addition to giving a new characterization of Kolmogorov complexity, Theorem 6.1 enables us to derive bounds on dimension from known bounds on Kolmogorov complexity. For example, we have the following. (Note: $\dim(|w|)$ is $\dim(z)$, where $z$ is the $|w|^{\text{th}}$ string in a standard enumeration of $\{0,1\}^*$.)

**Corollary 6.2.** *There exist constants $c_1, c_2 \in \mathbb{N}$ with the following properties.*

1. *For all $w \in \{0,1\}^*$, $\dim(w) \leq 1 + \dim(|w|) + \frac{c_1}{|w|}$.*

2. *For all $n, r \in \mathbb{N}$, if we choose $w \in \{0,1\}^n$ according to the uniform probability measure on $\{0,1\}^n$, then*
$$\Pr\left[\dim(w) > 1 + \dim(|w|) - \frac{r}{|w|}\right] > 1 - 2^{c_2 - r}.$$

*Proof.* This follows immediately from Theorem 6.1 and Theorem 3.3.1 of [22]. □

Thus most strings $w$ satisfy
$$\dim(w) = 1 + \dim(|w|) \pm O\left(\frac{1}{|w|}\right).$$

It follows readily that $\dim(w)$ often exceeds 1.

Theorem 6.1 characterizes Kolmogorov complexity in terms of dimension. We use this to give a new proof of a recent characterization of the dimensions of sequences.

**Corollary 6.3.** (Mayordomo [28]) *For all $S \in \mathbf{C}$,*
$$\dim(S) = \liminf_{n \to \infty} \frac{K(S[0..n-1])}{n}.$$

*Proof.* By Theorem 6.1,
$$\liminf_{n \to \infty} \frac{K(S[0..n-1])}{n} = \liminf_{n \to \infty} \dim(S[0..n-1]),$$
so the corollary follows by Theorem 5.9. □

We conclude this section with an easy proof that there exist sequences of all dimensions in [0,1].

**Theorem 6.4.** *For every $\alpha \in [0,1]$, $\mathrm{DIM}_\alpha \neq \emptyset$.*

*Proof.* Let $\alpha \in [0,1]$. If $\alpha = 0$ or $\alpha = 1$, then $\mathrm{DIM}_\alpha \neq \emptyset$ by Theorem 4.6 or Observation 4.4, respectively, so assume that $\alpha \in (0,1)$. Let $R \in \mathrm{RAND}$, and let $S$ be the sequence constructed by the following nonterminating, noncomputable procedure.

    **for** $n := 0$ **to** $\infty$ **do**
    $S[n] := $ **if** $K(S[0..n-1]) \leq \alpha n$ **then** $R[n]$ **else** $0$

Every sequence of the from $R' = wR[|w|..\infty)$ is random and thus satisfies $\frac{1}{n}K(R'[0..n-1]) \to 1$ as $n \to \infty$ [22]. On the other hand, every sequence of the form $T = w0^\infty$ satisfies $K(T[0..n-1]) = o(n)$ as $n \to \infty$ [22]. Finally, it is well-known that there is a constant $c \in \mathbb{N}$ such that for all $w \in \{0,1\}^*$ and $b \in \mathbb{N}$, $|K(wb) - K(w)| \leq c$. These three things together imply that the sequence $S$ satisfies
$$\lim_{n \to \infty} \frac{K(S[0..n-1])}{n} = \alpha,$$
whence $S \in \mathrm{DIM}_\alpha$ by Corollary 6.3. □

**Corollary 6.5.** *For every $\alpha \in [0,1]$,*

$$\mathrm{cdim}(\mathrm{DIM}_\alpha) = \alpha.$$

*Proof.* This follows immediately from Theorems 4.1 and 6.4. □

We conclude this section by discussing some earlier work relating martingales, supermartingales, and Kolmogorov complexity to Hausdorff dimension. Schnorr [37, 39] defined a martingale $d$ to have *exponential order* on a sequence $S$ if

$$\limsup_{n \to \infty} \frac{\log d(S[0..n-1])}{n} > 0 \tag{6.7}$$

and proved that no computable martingale can have exponential order on a Church-stochastic sequence. Terwijn [48] has noted that (6.7) is equivalent to the existence of an $s < 1$ for which the $s$-gale $d^{(s)}(w) = 2^{(s-1)|w|}d(w)$ succeeds on $S$. Thus, in the terminology of [25], Schnorr's result says that the set $\{S\}$ has computable dimension 1 for every Church-stochastic sequence $S$.

Ryabko [32] proved that

$$\dim_\mathrm{H}\left(\left\{S \,\bigg|\, \liminf_{n\to\infty} \frac{K(S[0..n-1])}{n} \leq \alpha\right\}\right) = \alpha, \tag{6.8}$$

and Cai and Hartmanis [3] proved that

$$\dim_\mathrm{H}\left(\left\{S \,\bigg|\, \liminf_{n\to\infty} \frac{K(S[0..n-1])}{n} = \alpha\right\}\right) = \alpha \tag{6.9}$$

for all $\alpha \in [0,1]$. In light of Corollary 6.3, (6.8) and (6.9) say that $\dim_\mathrm{H}(\mathrm{DIM}_{\leq \alpha}) = \alpha$ and $\dim_\mathrm{H}(\mathrm{DIM}_\alpha) = \alpha$, so (6.8) and (6.9) can be regarded as classical analogs of Theorem 4.7(1) and Corollary 6.5, respectively.

Ryabko [33] proved that

$$\dim_\mathrm{H}(X) \leq \sup\left\{\liminf_{n\to\infty} \frac{K(S[0..n-1])}{n} \,\bigg|\, S \in X\right\} \tag{6.10}$$

for all $X \subseteq \mathbf{C}$, and Staiger [46] established the existence of sets $X \subseteq \mathbf{C}$ for which

$$\dim_\mathrm{H}(X) < \sup\left\{\limsup_{n\to\infty} \frac{K(S[0..n-1])}{n} \,\bigg|\, S \in X\right\}. \tag{6.11}$$

By Theorem 4.1 and Corollary 6.3, (6.10) can now be seen as a statement of Observation 3.7(2).

Ryabko [35] and Staiger [47] defined the *exponent of increase* of a martingale $d$ on a sequence $S$ to be the number

$$\lambda_d(S) = \limsup_{n\to\infty} \frac{\log d(S[0..n-1])}{n}, \tag{6.12}$$

which is the left-hand side of (6.7). (We are using Staiger's notation here.) Both papers paid particular attention to the quantity

$$\lambda(S) = \sup\{\lambda_d(S) | d \text{ is a computable martingale}\}. \tag{6.13}$$

By Terwijn's above-mentioned observation, $1 - \lambda(S)$ is precisely the computable dimension of $\{S\}$ in the terminology of [25]. Ryabko [35] proved that

$$\lambda(S) \leq 1 - \liminf_{n \to \infty} \frac{K(S[0..n-1])}{n} \tag{6.14}$$

for every sequence $S$. By Corollary 6.3, we can now regard (6.14) as stating that $\dim(S)$ is no greater than the computable dimension of $\{S\}$. Ryabko [35] also proved that

$$\dim_H(\{S | \lambda(S) \geq \alpha\}) = 1 - \alpha \tag{6.15}$$

for all $\alpha \in [0,1]$. This is yet another analog of Theorem 4.7(1), saying that for all $\alpha \in [0,1]$ the set $\text{DIM}_\alpha^{\text{comp}}$, consisting of all sequences $S$ such that the computable dimension of $\{S\}$ is at most $\alpha$, has Hausdorff dimension $\alpha$. (Note: The earlier paper [34] proved results similar to (6.14) and (6.15), but with $\lambda(S)$ replaced by a different quantity, which we may call $\lambda'(S)$, in which the algorithm for the martingale is only required to halt on inputs of the form $w$, $w0$, or $w1$ for prefixes $w$ of $S$. It is easy to see that $\lambda'(S)$ is bounded below by $\lambda(S)$ and above by $1 - \dim(S)$.)

Staiger [47] provided even more insights. If $\mathbf{d} = \mathbf{d}^{(1)}$ is the optimal constructive supermartingale of Theorem 3.6 above, then Staiger's $\lambda_\mathbf{d}(S)$ is exactly $1 - \dim(S)$. He proved that

$$\dim_H(X) = \sup \left\{ \liminf_{n \to \infty} \frac{K(S[0..n-1])}{n} \,\middle|\, S \in X \right\} \tag{6.16}$$

for every $\Sigma_2^0$ set $X \subseteq \mathbf{C}$. In light of Theorem 4.1 and Corollary 6.3, this is equivalent to the result by Hitchcock [14], mentioned in section 4 above, that $\text{cdim}(X) = \dim_H(X)$ for every $\Sigma_2^0$ set $X \subseteq \mathbf{C}$. (It should be noted however, that the Staiger and Hitchcock results both preceded the Mayordomo [28] proof of Corollary 6.3 and that Hitchcock's result holds for arbitrary unions of $\Pi_1^0$ sets.) Staiger [47] also proved that

$$\sup \left\{ \inf_{S \in X} \lambda_d(S) \,\middle|\, d \text{ is a computable martingale} \right\} = 1 - \dim_H(X) \tag{6.17}$$

for every $\Sigma_2^0$ set $X \subseteq \mathbf{C}$. It is now easy to see that this is equivalent to the result by Hitchcock [14] that the computable dimension of a $\Sigma_2^0$ set $X \subseteq \mathbf{C}$ is precisely its Hausdorff dimension. Finally, Staiger [47] characterized Hausdorff dimension in terms of entropy rates, and Staiger [45] gave an enjoyable exposition of his and Ryabko's results in terms of an infinite game.

This brief review does not exhaust the results of the cited papers, but it does indicate the emergence of a rich network of relationships among martingales, supermartingales, Kolmogorov complexity, Hausdorff dimension, constructive dimension, and computable dimension.

## 7 Dimension and Biased Randomness

We now investigate the dimensions of sequences that are random relative to computable sequences of convergent biases. We first recall two known theorems concerning such sequences.

Given a bias sequence $\vec{\beta} = (\beta_0, \beta_1, \beta_2, \ldots)$, we write $\text{RAND}_{\vec{\beta}}$ for the set of all sequences that are random relative to the $\vec{\beta}$-coin-toss probability measure $\mu^{\vec{\beta}}$ defined in section 2. For each nonempty string $w \in \{0,1\}^+$, let

$$\text{freq}(w) = \frac{\#(1, w)}{|w|},$$

where $\#(b, w)$ is the number of times the bit $b$ occurs in $w$. For each $\beta \in [0, 1]$, we define the set

$$\text{FREQ}_\beta = \left\{ S \in \mathbf{C} \,\Big|\, \lim_{n \to \infty} \text{freq}(S[0..n-1]) = \beta \right\}.$$

The following well-known theorem is a constructive version of the strong law of large numbers.

**Theorem 7.1.** (folklore) *If $\vec{\beta}$ is a computable sequence of biases that converge to $\beta \in [0, 1]$, then $\text{RAND}_{\vec{\beta}} \subseteq \text{FREQ}_\beta$.*

**Definition.** Two sequences of biases $\vec{\beta}$ and $\vec{\beta}'$ are *square-summably equivalent*, and we write $\vec{\beta} \approx^2 \vec{\beta}'$, if $\sum_{i=0}^{\infty} (\beta_i - \beta'_i)^2 < \infty$.

The next theorem is a constructive version of a classical theorem of Kakutani [15].

**Theorem 7.2.** (van Lambalgen [49, 50], Vovk [51]) *Let $\vec{\beta}$ and $\vec{\beta}'$ be computable sequences of biases that converge to $\beta \in (0, 1)$.*

1. *If $\vec{\beta} \approx^2 \vec{\beta}'$, then $\text{RAND}_{\vec{\beta}} = \text{RAND}_{\vec{\beta}'}$.*

2. *If $\vec{\beta} \not\approx^2 \vec{\beta}'$, then $\text{RAND}_{\vec{\beta}} \cap \text{RAND}_{\vec{\beta}'} = \emptyset$.*

It is well-known (and easy to see) that a real number is $\Delta_2^0$-computable if and only if it is the limit of a computable sequence of reals. Thus Theorems 7.1 and 7.2 tell us that for each $\Delta_2^0$-computable bias $\beta \in (0, 1)$, the set $\text{FREQ}_\beta$ contains infinitely many disjoint sets of the form $\text{RAND}_{\vec{\beta}}$, where $\vec{\beta}$ is a computable sequences of biases converging to $\beta$. This section is concerned with the dimensions of the sequences in these sets $\text{RAND}_{\vec{\beta}}$. Our main result uses three lemmas.

Our first lemma follows immediately from a result in [25], but it is central to our development and a direct proof is brief, so we give it here. Recall the notation $\text{E}_\beta g(\vec{\alpha})$, the binary entropy function $\mathcal{H}(\beta)$, and the binary Kullback-Leibler divergence $D(\beta \,\|\, \alpha)$ discussed in section 2.

**Lemma 7.3.** *For all $\beta \in [0, 1]$, $\text{cdim}(\text{FREQ}_\beta) \leq \mathcal{H}(\beta)$.*

*Proof.* Let $\beta \in [0, 1]$, and let $s$ be a rational number with $s > \mathcal{H}(\beta)$. It suffices to show that $\text{cdim}(\text{FREQ}_\beta) \leq s$.

Let $\epsilon = \frac{s - \mathcal{H}(\beta)}{4}$. Fix a rational number $r \in (0, 1)$ such that

$$D(\beta \,\|\, r) < \epsilon. \tag{7.1}$$

Define $d : \{0, 1\}^* \to [0, \infty)$ by the recursion

$$d(\lambda) = 1,$$

$$d(w0) = 2^s (1 - r) d(w),$$

$$d(w1) = 2^s r d(w).$$

It is clear that $d$ is a constructive $s$-gale.

To see that $\text{FREQ}_\beta \subseteq S^\infty[d]$, let $S \in \text{FREQ}_\beta$. For all $n \in \mathbb{Z}^+$, let $w_n = S[0..n-1]$ and $\rho_n = \text{freq}(w_n)$. Since $S \in \text{FREQ}_\beta$, there exists $n_0 \in \mathbb{Z}^+$ such that for all $n \geq n_0$,

$$\mathcal{H}(\rho_n) < \mathcal{H}(\beta) + \epsilon \tag{7.2}$$

and
$$D(\rho_n \parallel r) < D(\beta \parallel r) + \epsilon. \tag{7.3}$$
For all $n \in \mathbb{Z}^+$ we have
$$d(w_n) = 2^{sn} r^{\#(1,w_n)} (1-r)^{\#(0,w_n)},$$
so
$$\begin{aligned}
\log d(w_n) &= n[s + \rho_n \log r + (1-\rho_n) \log(1-r)] \\
&= n\left[s - \mathrm{E}_{\rho_n} \log \frac{1}{r}\right] \\
&= n\left[s - \mathrm{E}_{\rho_n} \log \left(\frac{1}{\rho_n} \cdot \frac{\rho_n}{r}\right)\right] \\
&= n\left[s - \mathcal{H}(\rho_n) - D(\rho_n \parallel r)\right].
\end{aligned}$$

It follows by (7.1), (7.2), and (7.3) that for all $n \geq n_0$,
$$\log d(w_n) > n[s - \mathcal{H}(\beta) - 3\epsilon] = \epsilon n.$$
Thus $S \in S^\infty[d]$. This shows that $\mathrm{FREQ}_\beta \subseteq S^\infty[d]$, whence $\mathrm{cdim}(\mathrm{FREQ}_\beta) \leq s$. □

Our second lemma gives an asymptotic estimate of $\log \mu^{\vec{\beta}}(S[0..n-1])$ when $\vec{\beta}$ converges to $\beta \in (0,1)$ and $S$ has limiting frequency $\beta$.

**Lemma 7.4.** *If $\vec{\beta}$ is a bias sequence that converges to $\beta \in (0,1)$, then for all $S \in \mathrm{FREQ}_\beta$,*
$$\log \mu^{\vec{\beta}}(S[0..n-1]) = -\mathcal{H}(\beta)n + o(n)$$
*as $n \to \infty$.*

*Proof.* Using the abbreviations
$$\begin{aligned}
\tau_i &= \begin{cases} \log(1-\beta_i) & \text{if } S[i] = 0 \\ \log \beta_i & \text{if } S[i] = 1, \end{cases} \\
\bar{\tau}_i &= \begin{cases} \log(1-\beta) & \text{if } S[i] = 0 \\ \log(\beta) & \text{if } S[i] = 1, \end{cases} \\
\alpha_n &= \mathrm{freq}(S[0..n-1]),
\end{aligned}$$
the hypothesis tells us that
$$\begin{aligned}
\log \mu^{\vec{\beta}}(S[0..n-1]) &= \sum_{i=0}^{n-1} \tau_i \\
&= \sum_{i=0}^{n-1} (\bar{\tau}_i + o(1)) \\
&= \left(\sum_{i=0}^{n-1} \bar{\tau}_i\right) + o(n) \\
&= n[(1-\alpha_n)\log(1-\beta) + \alpha_n \log \beta] + o(n) \\
&= n[(1-\beta-o(1))\log(1-\beta) + (\beta+o(1))\log \beta] + o(n) \\
&= -\mathcal{H}(\beta)n + o(n)
\end{aligned}$$
as $n \to \infty$. □

Our third lemma is the crucial one. Its brief proof uses a natural transformation of an $s$-supergale to a $\vec{\beta}$-supermartingale.

**Lemma 7.5.** *If $\vec{\beta}$ is a computable sequence of biases that converge to $\beta \in (0,1)$, then for every computable $s \in [0, \mathcal{H}(\beta))$ and every constructive $s$-supergale $d$, the set $S^\infty[d]$ has constructive $\vec{\beta}$-measure 0.*

*Proof.* Let $\vec{\beta}$, $\beta$, $s$, and $d$ be as given. By Theorem 7.1, the set $\text{FREQ}_\beta$ has constructive $\vec{\beta}$-measure 1, so it suffices to show that the set $S^\infty[d] \cap \text{FREQ}_\beta$ has constructive $\vec{\beta}$-measure 0. Let

$$\sigma(w) = 2^{-s|w|} d(w)$$

for all $w \in \{0,1\}^*$. By Observation 3.2, $\sigma$ is a 0-supergale. Since $d$ is constructive and $s$ and $\vec{\beta}$ are computable it follows that the function

$$d' = \frac{\sigma}{\mu^{\vec{\beta}}}$$

is a constructive $\vec{\beta}$-supermartingale.

Now let $S \in \text{FREQ}_\beta$, and for each $n \in \mathbb{N}$, let $w_n = S[0..n-1]$. Since $s < \mathcal{H}(\beta)$, Lemma 7.4 tells us that for sufficiently large $n \in \mathbb{N}$,

$$sn + \log \mu^{\vec{\beta}}(w_n) < 0,$$

whence

$$d'(w_n) = \frac{d(w_n)}{2^{sn} \mu^{\vec{\beta}}(w_n)} > d(w_n).$$

This shows that $S^\infty[d] \cap \text{FREQ}_\beta \subseteq S^\infty[d']$. Thus $d'$ testifies that $S^\infty[d] \cap \text{FREQ}_\beta$ has constructive $\vec{\beta}$-measure 0. □

By Lemma 7.3 every sequence in $\text{FREQ}_\beta$ has dimension at most $\mathcal{H}(\beta)$. This upper bound is not in general tight. For example, if $\beta$ is $\Delta_2^0$-computable, it is easy to see that there are sequences of dimension 0 in $\text{FREQ}_\beta$. Nevertheless, the following theorem says that the upper bound $\mathcal{H}(\beta)$ is achieved by every sequence in each of the sets $\text{RAND}_{\vec{\beta}}$ for which $\vec{\beta}$ is computable and converges to $\beta$.

**Theorem 7.6.** *If $\vec{\beta}$ is a computable sequence of biases that converge to $\beta \in (0,1)$ and $R \in \text{RAND}_{\vec{\beta}}$, then $\dim(R) = \mathcal{H}(\beta)$.*

*Proof.* Assume the hypothesis. By Theorem 7.1 and Lemma 7.3, $\dim(R) \leq \mathcal{H}(\beta)$. To see that $\dim(R) \geq \mathcal{H}(\beta)$, let $s \in [0, \mathcal{H}(\beta))$, and let $d$ be a constructive $s$-supergale. By Lemma 7.5, $S^\infty[d]$ has constructive $\vec{\beta}$-measure 0. Since $R \in \text{RAND}_{\vec{\beta}}$, this implies that $R \notin S^\infty[d]$. Since this holds for all $s \in [0, \mathcal{H}(\beta))$ and all constructive $s$-supergales $d$, it follows that $\dim(R) \geq \mathcal{H}(\beta)$. □

Note that Observation 4.4 is exactly the case $\vec{\beta} = (\frac{1}{2}, \frac{1}{2}, \frac{1}{2}, \ldots)$ of Theorem 7.6. Note also that Theorem 7.6 can be used to give a second (albeit less informative) proof of Theorem 4.5.

Besicovitch [1] proved that $\dim_H(\text{FREQ}_{\leq \beta}) = \mathcal{H}(\beta)$ for all $\beta \in [0, \frac{1}{2}]$, where

$$\text{FREQ}_{\leq \beta} = \left\{ S \in \mathbf{C} \,\Big|\, \limsup_{n \to \infty} \text{freq}(S[0..n-1]) \leq \beta \right\}.$$

Good [12] conjectured that the limit supremum could be replaced by a limit here, thus obtaining $\dim_H(\text{FREQ}_\beta) = \mathcal{H}(\beta)$ for all $\beta \in [0,1]$. Eggleston [9] (see also [2, 11]) proved Good's conjecture. The following corollary is a constructive version of Eggleston's theorem.

**Corollary 7.7.** *If $\beta \in [0,1]$ is $\Delta_2^0$-computable, then* $\text{cdim}(\text{FREQ}_\beta) = \mathcal{H}(\beta)$.

*Proof.* If $\beta = 0$ or $\beta = 1$, this follows immediately from Lemma 7.3, so assume that $\beta \in (0,1)$ is $\Delta_2^0$-computable. Then there is a computable bias sequence $\vec{\beta}$ that converges to $\beta$, so Lemma 7.3 and Theorems 7.1 and 7.6 tell us that $\text{cdim}(\text{FREQ}_\beta) = \mathcal{H}(\beta)$. □

Computable bias sequences that converge slowly to $\frac{1}{2}$ have played an important role in the investigation of stochasticity versus randomness. First, Theorem 7.2 implies that if $\vec{\beta}$ is a bias sequence such that $\sum_{i=0}^\infty (\beta_i - \frac{1}{2})^2 = \infty$, then $\text{RAND}_{\vec{\beta}} \cap \text{RAND} = \emptyset$. Also, van Lambalgen [49, 50] proved that if $\vec{\beta}$ is any computable bias sequence that converges to $\frac{1}{2}$, then every element of $\text{RAND}_{\vec{\beta}}$ is Church-stochastic. Taking $\vec{\beta}$ to converge to $\frac{1}{2}$, but to do so slowly enough that $\sum_{i=0}^\infty (\beta_i - \frac{1}{2})^2 = \infty$ (e.g., $\beta_i = \frac{1}{2} + \frac{1}{\sqrt{i+4}}$), this gave a new proof that not every Church-stochastic sequence is random. More significantly, Shen' [41] strengthened van Lambalgen's latter result by showing that if $\vec{\beta}$ is any computable bias sequence that converges to $\frac{1}{2}$, then every element of $\text{RAND}_{\vec{\beta}}$ is Kolmogorov-Loveland stochastic. Again taking $\vec{\beta}$ to converge to $\frac{1}{2}$ slowly enough that $\sum_{i=0}^\infty (\beta_i - \frac{1}{2})^2 = \infty$, this allowed Shen' to conclude that not every Kolmogorov-Loveland stochastic sequence is random, thereby solving a twenty-year-old problem of Kolmogorov [16, 18] and Loveland [23, 24]. Theorems 7.6 and 7.2 have the following immediate consequence concerning such sequences $\vec{\beta}$.

**Corollary 7.8.** *If $\vec{\beta}$ is a computable sequence of biases that converge to $\frac{1}{2}$ slowly enough that $\sum_{i=0}^\infty (\beta_i - \frac{1}{2})^2 = \infty$, then*

$$\text{RAND}_{\vec{\beta}} \subseteq \text{DIM}_1 - \text{RAND}.$$

That is, every sequence that is random with respect to such a bias sequence $\vec{\beta}$ is an example of a sequence that has dimension 1 but is not random.

**Acknowledgments.** I thank Elvira Mayordomo, John Hitchcock, and Bas Terwijn for very helpful discussions. I also thank Ludwig Staiger for several useful observations on a prior draft of this paper (especially for noting my flawed use of gales in formulating constructive dimension) and for extending my knowledge of his and Ryabko's earlier work on Hausdorff dimension, Kolmogorov complexity, and martingales.